\documentclass[10pt,prl,twocolumn] {revtex4}

\usepackage[dvips]{graphicx}
\begin{document}

\title{Life time of superconductive state in long overlap Josephson junction}

\author{\firstname{K.G.} \surname{Fedorov}}
\author{\firstname{A.L.} \surname{Pankratov}}
\affiliation{Institute for Physics of Microstructures of RAS,
GSP-105, Nizhny Novgorod, 603950, Russia}
\email[]{alp@ipm.sci-nnov.ru}

\begin{abstract}
The computer simulations of fluctuational dynamics of the long
overlap Josephson junction in the frame of the sine-Gordon model
with a white noise source have been performed. It has been
demonstrated that for the case of constant critical current density
the mean life time (MLT) of superconductive state increases with
increasing the junction's length and for homogeneous bias current
distribution MLT tends to a constant, while for inhomogeneous
current distribution MLT quickly decreases after approaching of a
few Josephson lengths. The mean voltage versus junction length
behaves inversely in comparison with MLT.
\end{abstract}
\maketitle

At the present time in the area of quantum calculations serious
hopes are connected with the possibility to create qubits on the
basis of point \cite{qubit}, and distributed Josephson junctions
\cite{WKU}. The main advantage of qubits based on Josephson
junctions is the relative simplicity of manufacturing and
integrating in one circuit in comparison with other possible
realizations of qubits \cite{O-q}. It is necessary to point, that at
the present time all Josephson junctions are manufactured with the
use of electron-beam lithography \cite{D-arq},\cite{qubit} and can
always be considered as distributed. One of the most fundamental
problems of all quantum calculations is connected with the
destruction of entangled states of qubits through the interaction
with the environment. This event is characterized by the time of
decoherence, the maximal increase of which is needed for the design
of quantum computers. At the same time, due to mathematical
difficulties, even in the case of a long Josephson junction in the
presence of thermal fluctuations, the life time of superconductive
state is not studied well enough (see, e.g., \cite{BL}-\cite{Kusm}).
Mostly, annular and infinitely long junctions have been considered,
but it has been understood, that for short junctions the escape of
the whole string over the potential barrier occurs, while for long
junctions the creation of kink-antikink pairs is the main mechanism
of the escape process. In our recent works, dedicated to
investigation of noise-induced errors in Josephson junctions
\cite{PRL},\cite{APL}, only the model of point junction was
considered, so no dependences between fluctuational properties and
junction's length were studied. Therefore, at the present time an
important problem of investigation of superconductive state life
time with the aim to improve noise immunity of long Josephson
junctions is not yet solved. In this Letter we present results of
numerical investigation of superconductive state life time based on
the computer simulation of the sine-Gordon equation with white noise
source.

In the frame of the resistive McCumber-Stewart model \cite{Likh} the
phase difference of the order parameter $\varphi(x,t)$ of long
Josephson junction in the overlap geometry (see Fig. 1) is described
by the sine-Gordon equation:
\begin{eqnarray}
\beta\frac{\partial^2\varphi}{\partial t^2}+\frac{\partial\varphi}
{\partial t}-\frac{\partial^2\varphi}{\partial x^2}=i(x)-\sin
(\varphi)+i_f(x,t), \label{PSGE}
\end{eqnarray}
\begin{figure}[t]
\begin{picture}(200,150)(10,10)
\centering\includegraphics[width=8cm,height=6cm]{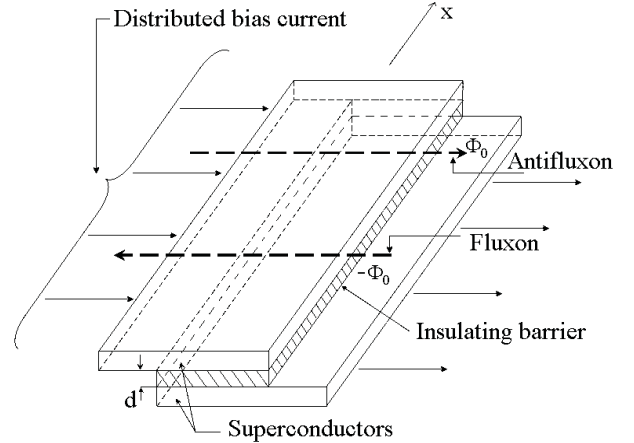}
\end{picture}
\parbox{240bp}
{\caption{The structure of distributed Josephson junction of the
"overlap" geometry.} \label{fig1}}
\end{figure}
with the following boundary conditions:
\begin{equation}
\frac{\partial\varphi(0,t)}{\partial x}=
\frac{\partial\varphi(L,t)}{\partial x}=\Gamma. \label{Bndr}
\end{equation}
Here the time and the space are normalized to the inverse
characteristic frequency $\omega_{c}^{-1}$ and the Josephson length
$\lambda _{J}$, respectively,  $\beta=1/\alpha^2$ is the
McCumber-Stewart parameter, $\alpha$ is the damping, $i(x)$ is the
bias current density, normalized to the critical current density of
the junction, $i_{f}(x,t)$ is the fluctuational current density,
$\Gamma$ is the normalized magnetic field, $L=l/\lambda_J$ is the
dimensionless length of the junction. In the case where the
fluctuations are treated as white Gaussian noise with zero mean, and
the critical current density is fixed, its correlation function is:
\begin{equation}
\left<i_f(x,t)i_f(x',t')\right>=2\gamma \delta (x-x^{\prime})\delta
(t-t^{\prime}), \label{CorrFunc2}
\end{equation}
where
\begin{equation}
\gamma = I_{T} / (J_{c}\lambda_J) \label{gam2}
\end{equation}
is the dimensionless noise intensity, $J_{c}$ is the critical
current density of the junction, $I_{T}=2ekT/\hbar$ is the thermal
current, $e$ is the electron charge, $\hbar$ is the Planck constant,
$k$ is the Boltzmann constant and $T$ is the temperature. It should
be noted, that in \cite{kautz} the total critical current
$I_c=\int_0^l J_c(x)dx$ (here $l=L\lambda_J$) was supposed to be
fixed, and the noise intensity in that case was proportional to the
dimensionless length of the junction $L$. In the frame of the
present paper we consider more interesting (from the practical point
of view) case where the critical current density is held constant
and the critical current increases with the junction's length.

Numerical solution of the sine-Gordon equation (\ref{PSGE}) with
boundary conditions (\ref{Bndr}) has been performed on the basis of
implicit finite-difference scheme \cite{Zhang} with the account of
the white noise source. Typical values of discretization steps are
$\triangle x = \triangle t = 0.1 - 0.02$, number of realizations are
$R = 1000 - 10000$.

In the frame of the present paper we will consider and compare two
cases of bias current distribution: homogeneous $i(x)=i_0$ and
inhomogeneous one, characteristic for a superconductive thin film
\cite{Likh},\cite{SV}:
\begin{equation}
i(x)=\frac{i_0 L}{ \pi \sqrt{ x  (L - x)}}.
\label{inh-distr}
\end{equation}

For the junctions with small capacitance $\beta\ll 1$ (large damping
$\alpha\gg 1$), the mean life time (MLT) of superconductive state
is, physically, the mean time till the generation of noise-induced
voltage pulse. The MLT is defined as the mean time of phase
$\varphi$ existence in the considered interval \cite{MP},\cite{ACP},
where $W(\varphi,x,t)$ is the probability density, and $P(t)$ is the
probability that the phase is located in the initial potential well:
\begin{eqnarray}\label{Tay}
\tau=\displaystyle{\int_0^{+\infty} t w(t)dt} = \int_0^{+\infty} P(t)dt,
\phantom {aaaaaa}\\
w(t)=-\displaystyle{\frac{\partial P(t)}{\partial t}}, \,
P(t)=\frac{1}{L} \int_0^L \int_{-\pi}^{\pi} W(\varphi,x,t) d\varphi
dx. \nonumber
\end{eqnarray}

\begin{figure}[h]
\begin{picture}(200,150)(10,10)
\centering\includegraphics[width=8cm,height=6cm]{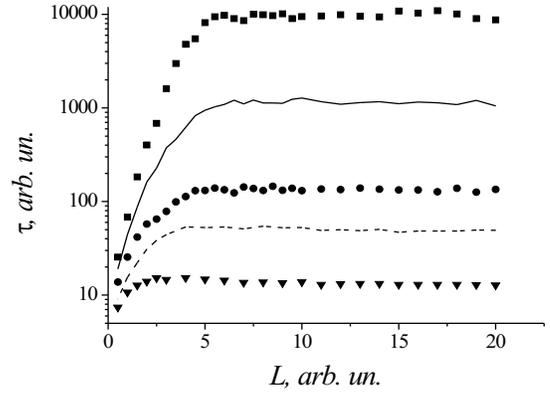}
\end{picture}
\parbox{240bp}
{\caption{The MLT of superconductive state of long Josephson
junction for the case of homogeneous bias current distribution with
$\Gamma = 0$, $\beta=0.01$: squares - $i_0=0.5$, $\gamma=0.3$;
circles - $i_0=0.7$, $\gamma=0.3$; triangles - $i_0=0.9$,
$\gamma=0.3$; solid line - $\gamma=0.2$, $i_0=0.7$; dashed line -
$\gamma=0.4$, $i_0=0.7$.} \label{fig2}}
\end{figure}

\begin{figure}[h]
\begin{picture}(200,150)(10,10)
\centering\includegraphics[width=8cm,height=6cm]{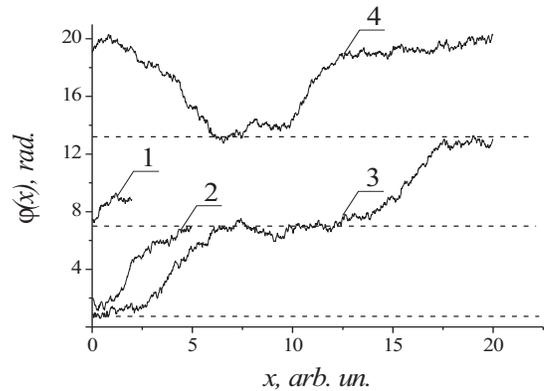}
\end{picture}
\parbox{240bp}
{\caption{Phase evolution in the long Josephson junction depending
on its length: 1 - $L=2$, 2 - $L=5$,  3 - $L=20$, $t=t_1$, 4 -
$L=20$, $t=t_2>t_1$.} \label{fig3}}
\end{figure}

First, let us consider the case of homogeneous bias current
distribution $i(x)=i_0$. The MLT of superconductive state versus
length of a Josephson junction, calculated numerically on the basis
of definition (\ref{Tay}) with the delta-shaped initial distribution
at $\varphi_0=\arcsin(i_0)$, is presented in Fig. 2. As it is seen,
the MLT of superconductive state for the case of constant critical
current density rises for small lengths $L\leq 1$ and reaches a
constant for $L \geq  5$. Such behavior is explained by the
existence of two different mechanisms of the string escape over the
potential barrier \cite{BL} (see also \cite{M88}-\cite{Kusm}): it
may escape either as a whole, or by forming kink-antikink pairs. For
$L\leq 1$ with the increase of the string length, by the interaction
of elementary parts of the string with their neighbors, the phase is
stronger kept in the initial potential well which leads to the
increase of the life time. Even if the kink-antikink pair is
fluctuationally formed, but the string is too short, the attraction
between the kink and antikink outweighs the driving force $i$ and
the incipient nucleus collapses returning the string to the initial
potential well. If, however, the string is long enough, then the
driving force predominates and the nucleus expands, speeding up the
escape process. These mechanisms are illustrated in Fig. \ref{fig3}:
while for $L < 1$ the first mechanism is predominated, for $L \geq
5$ the second one leads to the saturation of MLT. It is important to
mention, that for the considered overlap junction (see boundary
conditions (\ref{Bndr})) not only the creation of kink-antikink
pairs (curve 4), but also creation of one (curve 2), two (curve 3),
or even more kinks is possible. As it is seen from Fig \ref{fig2},
the constant of saturation depends on thermal noise intensity
$\gamma$ (decreases with the increase of $\gamma$), and bias current
density $i$ (also decreases with the increase of $i$, because of
decrease of the potential barrier height).
\begin{figure}[h]
\begin{picture}(200,150)(10,10)
\centering\includegraphics[width=8cm,height=6cm]{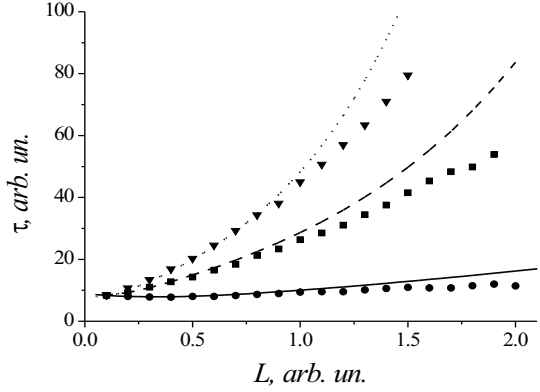}
\end{picture}
\parbox{240bp}
{\caption{Comparison between numerical simulations of MLT (symbols)
and formula (\ref{MPF}) (lines) for the case of homogeneous bias
current distribution with $\Gamma = 0$, $\beta=0.01$, $i_0=0.7$;
circles and solid line - $\gamma=1$, squares and dashed line -
$\gamma=0.3$, triangles and dotted line - $\gamma=0.2$.}
\label{fig4}}
\end{figure}

It is interesting to check the limiting transition of a long
junction to a point one for small junction lengths $L\ll 1$. For a
point junction in the case of large damping $\beta \ll 1$ the MLT of
superconductive state was found analytically \cite{MP} for arbitrary
value of noise intensity. That formula, however, was obtained for
the case of constant critical current $I_c$, and the noise intensity
in that case was: $\gamma_s=I_T/I_c$, where $I_c=J_c\lambda_J L$.
Substituting $\gamma_s$ as $\gamma_s=\gamma/L$ (where $\gamma$ is
given by (\ref{gam2})), one can get the following formula, that
should describe MLT of superconductive state for the case of a short
Josephson junction with constant critical current density:
\begin{eqnarray}
& \tau=\displaystyle{\frac{L}{\gamma}\left\{
\int_{\varphi_0}^{\varphi_2 }e^{-(\cos x +ix )L/\gamma }
\int_{\varphi_1}^xe^{(\cos \varphi +i\varphi )L/\gamma }
d\varphi dx+\right.} & \nonumber\\
& \displaystyle{\left.+\int_{\varphi_1}^{\varphi_2 }e^{(\cos \varphi
+i\varphi )L/\gamma }d\varphi\cdot \int_{\varphi_2 }^\infty
e^{-(\cos \varphi +i\varphi )L/\gamma }d\varphi \right\}}. &
\label{MPF}
\end{eqnarray}
where $\varphi_0$ - coordinate of initial delta-shaped distribution,
$\varphi_{1,2}$ - boundaries of the interval, which define the
initial potential well. Indeed, as it is seen from Fig. 4, the
results of numerical simulations of a long junction perfectly
coincide with the formula (\ref{MPF}) not only in the limiting case
$L\ll 1$, but even up to $L\sim 1$ and for $L>1$ the coincidence is
better for larger noise intensity.

\begin{figure}[h]
\begin{picture}(200,150)(10,10)
\centering\includegraphics[width=8cm,height=6cm]{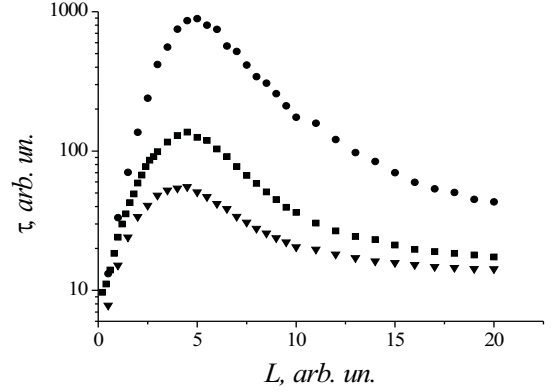}
\end{picture}
\parbox{240bp}
{\caption{The MLT of superconductive state of long Josephson
junction for the case of inhomogeneous critical current density with
$\Gamma = 0$, $\beta=0.01$, $\gamma=0.4$ and: circles - $i_{0}=0.5$,
squares - $i_{0}=0.6$, triangles - $i_{0}=0.7$.} \label{fig5}}
\end{figure}

In real situations it is hard to achieve the homogeneous bias
current distribution. Since a long overlap Josephson junction is
made on the basis of thin superconductive films, the bias current
distribution may be governed by the formula (\ref{inh-distr})
\cite{Likh},\cite{SV} for the case when the junction width is much
smaller than its length. Such bias current profile significantly
changes the picture of the life time of superconductive state
dependence on junctions's length. Results of study of the MLT in the
case of inhomogeneous bias current density are presented in Fig.
\ref{fig5}.

\begin{figure}[h]
\begin{picture}(200,150)(10,10)
\centering\includegraphics[width=8cm,height=6cm]{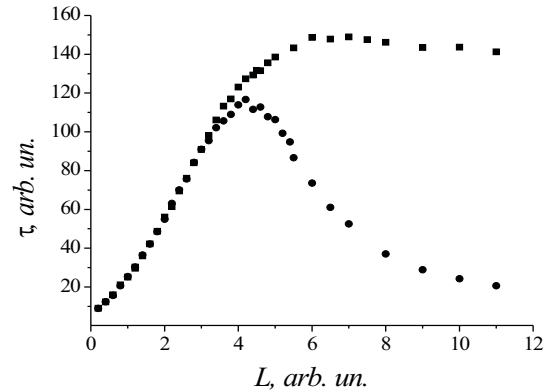}
\end{picture}
\parbox{240bp}
{\caption{Comparison of MLT of superconductive state of long
Josephson junction for different cases of bias current distribution
with $\Gamma = 0$, $\beta=0.01$, $\gamma=0.3$, $i_{0}=0.7$: circles
- inhomogeneous distribution, squares - homogeneous distribution.}
\label{fig6}}
\end{figure}

The main feature of such dependences is the existence of strong
maximum for the lengths $L\approx 5$. This effect can also be
explained by the existence of two mechanisms of the string escape.
It is obvious, that mechanism of MLT increase for small lengths is
absolutely the same as for the case of constant bias current
distribution (see Fig. \ref{fig6}). The main difference between the
cases of homogeneous and inhomogeneous bias current distribution is
the fact, that in inhomogeneous case there are unstable areas near
the edges of a junction ($x\approx0,L$) because there the bias
current is larger than the critical current according to the formula
(\ref{inh-distr}). These areas work as generators of "kink-antikink"
pairs. Due to the unstable character of a potential profile near the
edges, amount of those pairs is significantly larger than in the
junction with homogeneous bias current distribution. So, the
influence of the second mechanism on the life time for the
inhomogeneous bias current distribution is larger in comparison with
the homogeneous one. But the lengths of "switching on" of this
mechanism are the same as for the homogeneous case (see Fig.
\ref{fig6}).

On one hand, to experimentally observe the predicted above effect of
non-monotonous dependence of the MLT of superconductive state versus
junction's length, one needs to perform advanced measurements. On
the other hand, it is now believed \cite{kosh} that the use of more
homogeneous bias current distributions may increase the power and
decrease the linewidth of practical Flux-Flow Oscillators (FFOs),
based on long Josephson junctions. There are attempts to increase
the homogeneity by the increase either the junction's width or the
width of the idle regions (see \cite{kosh} for details) and the
check of the homogeneity of the current profile is only performed by
the measurements of the linewidth. This takes a long time and there
are many other different factors which affect the linewidth. So, a
reliable tool to detect a homogeneity of bias current distribution
in a particular FFO design is needed. To this end, the measurement
of a voltage-length characteristic for a fixed value of bias current
(slightly smaller than the critical current) can be recommended as
such a tool.
\begin{figure}[h]
\begin{picture}(200,150)(10,10)
\centering\includegraphics[width=8cm,height=6cm]{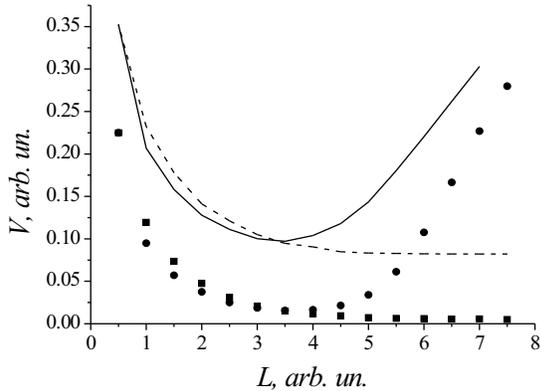}
\end{picture}
\parbox{240bp}
{\caption{Comparison of mean voltage $V(L)$ for the different cases
of bias current distribution with $i_{0}=0.9$, $\Gamma = 0$,
$\beta=0.01$; $\gamma = 0.05$: circles - inhomogeneous case, squares
- homogeneous case; $\gamma = 0.1$: solid line - inhomogeneous case,
dashed line - homogeneous case.} \label{fig7}}
\end{figure}

The mean noise-induced voltage versus junction's length demonstrates
the "inverse" behavior in comparison with the MLT (see Fig.
\ref{fig7}). For the case of constant current density and
homogeneous bias current distribution the mean voltage decreases for
small lengths $L\leq 1$ and reaches a constant for $L \geq  5$. For
the case of inhomogeneous current distribution the minimum of the
mean voltage versus length is observed. As one can see from Fig.
\ref{fig7}, the voltages for homogeneous and inhomogeneous cases
differ exponentially in the small noise limit for large lengths of
Josephson junctions. If, with increase of the junction's width, the
bias current distribution will be more and more close to the
homogeneous one, it can be clearly seen in $V(L)$ dependence.

In this work numerical and qualitative analysis of fluctuational
dynamics of long Josephson junctions is presented. It has been
demonstrated that for the case of constant critical current density
the mean life time (MLT) of superconductive state increases with
increasing the junctions's length and for homogeneous bias current
distribution MLT tends to a constant for $L
> 5$. However, for inhomogeneous current distribution MLT quickly
decreases after approaching a maximum for lengths around $L \approx
5$. Therefore, from the fluctuational stability point of view, there
is no reason to increase the length of a long Josephson junction
more than $L \approx 5$, excepting the cases, where increasing of
the junction length can improve other useful properties of Josephson
electronic devices. It has also been demonstrated, that the measure
of the homogeneity of bias current distribution may be performed on
the basis of voltage versus length characteristics for bias current,
smaller than the critical one: for junction lengths larger than
$5\lambda_J$, the voltages for bias current distribution
(\ref{inh-distr}) and the homogeneous one differ exponentially in
the limit of small noise intensity.

The work has been supported by ISTC (projects 2445 and 3174), and
the Russian state contract 02.442.11.7342.

\end{document}